
\magnification=1200
\newcount\eqnumber
\eqnumber=1
\def\eqnam#1{\xdef#1{\the\eqnumber}}
\def\neweq{\eqno(\the\eqnumber\global\advance\eqnumber by 1)}
\newcount\refnumber
\refnumber=1
\def\newref{\the\refnumber\global\advance\refnumber by 1}
\def\refname#1{\xdef#1{\the\refnumber}\newref}
\pageno=1
\rightline{BNL-49465}
\rightline{September 1993}
\vskip.5in
\centerline{\bf Lattice Gauge Theory -- Present Status}
\bigskip
\medskip
\centerline{\bf Michael Creutz}
\centerline{Physics Department, Brookhaven National Laboratory}
\centerline{Upton, NY 11973,USA}
\def\slash{/\kern-6pt}

\bigskip
\bigskip
\baselineskip=18pt

 \centerline {ABSTRACT}
\medskip
{\narrower
Lattice gauge theory is our primary tool for the study
of non-perturbative phenomena in hadronic physics.  In addition
to giving quantitative information on confinement, the approach is yielding
first principles calculations of hadronic spectra and matrix elements.
After years of confusion, there has been significant recent progress
in understanding issues of chiral symmetry on the lattice.
\par
}
\vfill
\noindent
Talk presented at HADRON 93, Como, Italy, June 1993.
\eject
Lattice gauge theory is a rather old subject, going back to Wilson's work
of the early 70's [\refname\wilsonref].  Through the
80's it grew into a major industry.  The field is currently dominated
by computer simulations, although it is in fact considerably
broader.  The main results are now presented annually at a
lattice conference attended by about
300 participants.  The proceedings of these meetings make a good source
of background material for the topic [\refname\latticeref].

The lattice program has rather grandiose goals: the first
principles solution of hadronic physics.  Indeed, sometimes the practitioners
get a bit overenthusiastic in stating what is possible.  On the other hand,
there is no other known way to obtain many of the quantities
currently being calculated.

Among these quantities are the hadronic spectra, the
hadronic matrix elements of operators of importance to weak decays, and
the properties of the quark gluon plasma.  Going beyond hadrons,
there has been extensive work on obtaining constraints on the Higgs particle.
There have also been simulations of curved space times, made with
the hope of getting a handle on quantum gravity.

So why do we put our field theory on a lattice?  From my
point of view, this is a mathematical trick.  The lattice provides an
ultraviolet cutoff which allows the system to be placed on a computer.
Perhaps most important, the cutoff is not based on perturbation theory, and
thus we can study non-perturbative physics, such as confinement.
Also, unlike with some other non-perturbative schemes, we have
a well defined system for study.

Wilson's formulation provides a rather elegant framework for these studies.
This begins with a discretization of the action
$$
S=\int d^4x F^{\mu \nu} F_{\mu\nu}\rightarrow {\beta\over 3} \sum_p {\rm
ReTr}(U_p) \eqnam{\actioneq}\neweq
$$
Here the sum is over all elementary squares or ``plaquettes'' of a
four dimensional simple hypercubic lattice.  The variable $U_p$
is an $SU(3)$ matrix which measures
the gauge field flux through the plaquette, and is the matrix
product of elementary
link variables surrounding the plaquette
$$
U_p= \prod _{\{i,j\}\in p} U_{i,j} \neweq
$$
where this product is understood in an ordered sense.
The individual link variables are identified
as the phase factors associated
with the gauge field along the respective link
$$
U_{i,j} \sim \exp(i \int_{x_i}^{x_j} A^\mu dx_\mu). \neweq
$$
The bare gauge coupling is related to the parameter $\beta$ in eq. (\actioneq)
$$
\beta={6 \over g_0^2} \neweq
$$

The phenomenon of asymptotic freedom relates the
coupling to the lattice spacing $a.$  In particular,
the bare coupling should go logarithmically to zero with the
scale on which it is defined.  Here this scale is the
lattice spacing; so, we have
$$
\alpha_0 \sim {1\over \beta_0 \ln (1/(a^2\Lambda^2))} \neweq
$$
where $\Lambda$ is an integration constant for the renormalization
group equation, and $\beta_0$ is a numerical constant.
Instead of regarding the coupling as a function
of the scale, we can invert this relation and consider the lattice spacing
as a function of the coupling
$$
a\sim {1\over\Lambda} \exp({-1\over2\beta_0\alpha_0}). \neweq
$$
If we now go to our lattice and measure
any dimensional quantity in lattice units, these relations give us a handle
on the
strong coupling constant.  Various physical
quantities can set the normalization.  In the earliest studies of
confinement it was usual to take the
string tension or Regge slope.  People interested in weak interaction
matrix elements often pick one of the
meson decay constants, such as $f_\pi$, to set the scale for other
measurements.  In a spectroscopic application one might pick a light
hadron mass, such as of the rho or nucleon.  For another specific
example, the Fermilab group recently studied the charmonium spectrum
on the lattice and extracted a value for the strong coupling constant
[\refname\fnalref].
Putting in corrections corresponding to four quark flavors, they quote
$$
\alpha_{\overline{MS}}(5{\rm Gev})=0.174\pm0.012 \neweq
$$
which is in reasonable agreement with experimental measurements.

Although the lattice represents a broad framework
for the nonperturbative definition of a field theory, the subject is
currently dominated by one approach, that of
Monte Carlo simulation.  This uses the analogy between the
Feynman path integral
$$
Z=\int dU e^{\beta/3  \sum {\rm Re Tr}(U_p)} \neweq
$$
and the ``partition function'' for a set of ``spins''
$\{U\}$ at ``temperature'' $1/\beta$.
This is easily simulated on a finite lattice by standard methods.
A Monte Carlo program sweeps over a lattice stored
in a computer memory and makes random changes biased by the above
``Boltzmann weight.''  The procedure generates a sequence of configurations
which mimic ``thermal equilibrium.''
$$
P(C)\sim  e^{\beta  \sum_p U_p} \neweq
$$

One of the captivating features of the technique is that the entire
lattice is available in the computer memory; so,
in principle one can measure anything.  On the other hand,
there are inherent
statistical fluctuations which may make some things hard to
extract.  This represents a new aspect of theoretical physics,
wherein theorists have statistical errors.
In addition these calculations have several sources of systematic errors,
such as effects of finite volume and finite lattice spacing.
Quark fields introduce further sources of error,
including extrapolations from heavy to physical quark masses.  Furthermore,
many calculations are made
feasible by what is termed the valence or quenched
approximation, wherein virtual
quark loops are neglected.

Indeed, quark fields introduce
serious unsolved problems.  These problems are by no means new,
having been with us since the beginnings of lattice gauge theory.
One problem is the issue of finding a reasonable
computer algorithm for simulating fermions.  In particular, since
the quark fields are anticommuting, the full action is not an
ordinary number, and the analogy with
classical statistical mechanics breaks down.  Algorithms in current
practice begin by formally integrating out the fermions to give a determinant
$$
\eqalign
{
Z&=\int dA\ d\bar\psi d\psi\ \exp(S_g + \bar \psi (\slash D +m) \psi)\cr
&=\int dA\ e^{S_g}\ \vert \slash D +m \vert. \cr
} \neweq
$$
This determinant is, however, of a rather huge matrix, and is
quite tedious to simulate.  Over the years
many clever tricks have been found to simplify the problem, but I regard
these approaches as still rather ugly.  In addition, if one wants to study
physics in a chemical potential, as would represent
background baryon density in a heavy ion experiment, then
no viable simulation algorithms are known.

The other old problem with quarks has to do with the issue of fermion
doubling and and chiral symmetry.  Here there has been
considerable recent progress, to which I will return later in this talk.

The difficulties with simulating dynamical fermions have led to the
majority of simulations being done in the
``valence'' or ``quenched'' approximation.  Here the feedback of the
fermion determinant on the dynamical gauge fields is ignored.  Hadrons
are studied via quark propagators in a gauge
field obtained in a simulation of gluon fields alone.  In terms of Feynman
diagrams, all gluonic exchanges are included between the constituent
quarks, but effects of virtual quark production, beyond simple
renormalization of the gauge coupling, are dropped.
The primary motivation is the saving of orders of magnitude in computer
time.  While this may seem a drastic approximation,
the fact that the naive quark model works so well
hints that things might not be so bad.

One of the longstanding goals of lattice calculations is an understanding of
hadronic spectra.  If we consider the correlation
between some operator $\phi$ taken at two widely separated points, we expect a
generic behavior
$$
\langle  \phi(x) \phi(0) \rangle  \longrightarrow e^{-Mx}  \neweq
$$
where $M$ is the mass of the lightest hadron which can be created by
$\phi$ acting on the vacuum.  Via such calculations using different operators,
the masses of a
large variety of states can be estimated.  In these calculations the
bare quark masses are parameters.  Lore based on chiral symmetry
suggests that as the quark masses go to zero, so will the
masses of the corresponding
pseudoscalar mesons, i.e. the pions.
Thus, the procedure is to adjust
the quark masses to get, say,
$m_\pi/m_\rho$ right, and
then all other mass ratios, such as $m_N/m_\rho$ should be determined.

An extensive recent calculation of this type was presented in
ref.~[\refname\weingartenref].  This particular project was carried out on a
specially built computer which ran for one year at an average speed
of six gigaflops.  These valence approximation results for the
light hadron masses are consistent
with experiment to within 6\%.

Another area of extensive investigation is the
phase transition of the vacuum to a plasma of free quarks and gluons
at a temperature of a few hundred MeV.  To study field theory at a finite
temperature, we use the fact that in a finite temporal box of length $t$
the path integral takes the form
$$
Z\sim \exp(-Ht) \neweq
$$
where $H$ is the quantum Hamiltonian.  This
is exactly the thermal partition function.  Both theoretical
analysis and numerical simulations have shown the existence of a
high temperature regime wherein
confinement is lost and
chiral symmetry is manifestly restored [\refname\gavairef].
The lattice Monte Carlo calculations have given us the best estimate
of the relevant transition temperature
$$\eqalign{
 &T_c\sim 235\ {\rm MeV,\ 0\ flavors}\cr
 &T_c\sim 150\ {\rm MeV,\ 2\ flavors.}\cr
  }\neweq
$$
The relatively large
difference between these numbers was somewhat unexpected.  Indeed,
this is the only place known where the valence approximation seems
to have a substantial physical effect.

A large subindustry in the lattice community is the evaluation
of hadronic matrix elements of operators relevant to processes
such as weak decays.  The standard electroweak theory makes precise
predictions for the relevant operators leading to weak decays, but
to relate these to observed decay rates requires the inclusion
of strong interaction initial and final state
corrections.  These are non-perturbative
in nature, and thus fall directly into the lattice gauge theorist's relm.
This is a rather large area of research which I cannot cover adequately here.
Instead, I defer to the recent review
of Bernard and Soni [\refname\bernardsoniref].  To quote one
recent result,
the decay constants for the $B$ and $D$ mesons have been obtained
in Ref.~[\refname\blsref].  Using
$f_\pi$ to set the overall scale, they find
$$\eqalign
{
&f_B=187(10)\pm34\pm15 MeV \cr
&f_{B_s}=207(9)\pm34\pm22 MeV\cr
&f_D=208(9)\pm35\pm12 MeV. \cr
}\neweq
$$

I now return to the problems with fermions and discuss some of
the issues concerning
chiral symmetry and fermion doubling.
It was realized quite early that if one naively discretizes the Dirac
equation on the lattice, one obtains extra particles.  Wilson showed
how these could be removed by adding the so called
``Wilson term'' which formally vanishes in the continuum limit.
Unfortunately, this term inherently violates chiral symmetry.
As many predictions have been based on this symmetry, the usual hope
is that in continuum limit it will return.  Meanwhile,
however, there is nothing special about massless quarks, and when the
cutoff is still in place one
must ``tune'' the bare parameters to make $m_\pi$ small.

It has recently been suggested that an infinite tower of heavy states may
solve this problem.  The basic mechanism is to absorb  the
extra species of the naive formulation into a band
of heavy states.  One formulation
of the idea was presented by Kaplan [\refname\kaplanref], and some
intriguing variations discussed by Frolov and Slavnov [\refname\slavnovref]
and by Neuberger and Narayanan [\refname\neubergerref].

Let me discuss the problem in somewhat more detail in
one space dimension.  A naive discretization of the Dirac Hamiltonian
is
$$
H_0=K\sum_j i(a_j^\dagger a_{j+1}-b_j^\dagger b_{j+1})
 + M\sum_j a^\dagger_j b_j + h.c. \neweq
$$
where $a_j$ and $b_j$ are fermionic annihilation operators on sites
$j$ located along a line.  They represent the upper and lower components
of a two component spinor.  The spectrum of single particle states
for this Hamiltonian is easily found in momentum space
$$
E^2=M^2+4K^2 \sin^2(q) \neweq
$$
where $q$ runs from 0 to $2\pi$.
Filling the negative energy states to form a Dirac sea, the physical
excitations consist of particles as well as
antiparticle ``holes.''  The doubling
problem is manifested in the fact that there are low energy excitations for
momenta $q$ in the vicinity of $\pi$ as well as 0.  In $D$ spatial dimensions,
this doubling increases to a factor of $2^D$.

A simple solution to the doubling was presented some time ago by
Wilson, who added a term that created a momentum dependent mass
$$
H=H_0-\ rK \sum_j (a_j^\dagger b_{j+1}+b_j^\dagger a_{j+1} + h.c.) \neweq
$$
where $r$ is called the
Wilson parameter.
The energy spectrum is now
$$
E^2=4K^2 \sin^2(q)+(M-2Kr\cos(q))^2 \neweq
$$
and we see that the states at $q$ near $\pi$ have a different
energy than those near $0$.  For the continuum limit, the parameters
should be adjusted so that the extra states become infinitely
heavy.

This Hamiltonian has a special behavior when $2Kr=M$.   In this case
one of the fermion species becomes massless.  This provides a mechanism
for obtaining light quarks and chiral symmetry.  Unfortunately, when
the gauge interactions are turned on, the parameters renormalize, and tuning
becomes necessary to maintain the massless quarks.
This is the basis of the conventional approach to chiral
symmetry with Wilson fermions; one tunes the ``hopping parameter'' $K$
until the pion is massless, and calls that the chiral limit.

The above Hamiltonian has some peculiar properties when we take
the hopping parameter larger than the critical value.  This region
has been discussed by Aoki and Gocksch [\refname\aokigockschref]
in the context of a spontaneous breaking of parity, although the connection
of their work to what I will discuss below is as yet obscure.

Restricting the above Hamiltonian to a finite box with
open boundaries, the states become discrete and fall into three
classes.  First there is a set of states with positive energy which
represent the particle band with energies above $\vert 2Kr-M \vert$.
Second, there are the complimentary negative energy states which
represent the Dirac sea.  Finally, if $K$ exceeds the critical
value $M/2r$, there are two isolated levels left near zero energy.
These levels are surface modes bound to the boundary of our finite box.
They are split
from zero energy by tunnelling from one boundary to the other; indeed,
they go to exactly zero energy in the infinite volume limit.

Kaplan [\kaplanref] has proposed to
use such zero modes on a boundary as the basis for a formulation
of chiral fermions.  The idea is to consider the above one dimensional
system as representing not a physical coordinate, but a hypothetical
``fifth'' dimension.  Our world, then, would be a four dimensional
interface representing a boundary in this extra dimension.  The physical
quarks and leptons are
the above surface modes.  Momentum in the
physical dimensions gives an additional contribution to their
energy, yielding a conventional relativistic spectrum.

The existence of the surface modes requires that the hopping parameter
exceeds a critical value.  A more general result
is that these modes exist whenever the hopping parameter
passes through the critical value as one passes through the surface.
I have simplified the discussion by having $K$ vanish outside
the supercritical region.

With the
physical transverse dimensions included, the critical value for the
hopping parameter in
the fifth dimension will depend
on the physical momenta.  By appropriately choosing the parameters,
only those states with low physical momentum will satisfy the
conditions for a surface mode to exist,
and all extra doubler modes can be eliminated
[\refname\jansenschmaltzref].  Thus the tuning
problem of the conventional approach is replaced with a large volume limit
for the new fifth dimension.

In this picture, the opposite walls have modes representing opposite
helicity.  Chiral anomalies appear through tunnelling between these
walls.  As discussed in Ref.~[\refname\callanharveyref], these anomalies
maintain their correct values even as the band of states in the extra
dimension goes to infinite mass.

The above discussion ignored the contributions of gauge fields.  To avoid
adding new unwanted degrees of freedom, it is most natural to place
gauge fields only on the links in the physical four dimensions of ordinary
space time.  It is perhaps simplest to follow Ref.~[\neubergerref]
and to think of the extra dimension
as representing a ``flavor'' space, with all the new flavors coupling
equivalently to the gauge fields.

One consequence of this picture is that
the same gauge field will couple to the surface modes on
both walls of a finite slab.  Thus in the simplest case both chiralities
of the fermions will be present, and one does not have an immediate lattice
formulation of the standard model of electroweak interactions.
At first sight this appears
discouraging; indeed, we know the
weak interactions require parity violation, and for conceptual reasons
we would like to have a non-perturbative formulation.

Formulating the standard model on the lattice presumably
requires some rather subtle features.  For example,
some time ago t'Hooft [\refname\thooftref] showed how instanton effects
give rise to a small baryon number violation in the standard model.
Any non-perturbative formulation must include such a phenomenon.  What
is exciting is that this extra dimension formulation may have a
mechanism to do just that.

In particular, the t'Hooft baryon violation is an
effect of anomalies, and anomalies in this surface picture are represented
by tunnelling through the extra dimension.  For this to work out,
baryon states on on one surface will need to mix with
lepton states on the other.  Then baryon violation could then arise as a
tunnelling between
these states through the ``fifth'' dimension.
The details of such a scheme still need to be
worked out; indeed, one must carefully assure that no anomalies
remain for the gauged currents.

To conclude this talk, let me make a few disconnected remarks.  As the lattice
provides a non-perturbative definition of a field theory, there have
been numerous efforts at using the methods on other models.  A particularly
active area has been towards understanding gravity.  The general idea
would be to discretize the points of space-time and then do
a sum over curvatures.  So far the results of this
program have been limited, but no other approach to quantum gravity
has yet been successful and the potential payoff is great.
A review of the subject was given by Jurkiewicz [\refname\gravityref].

Occasionally one hears suggestions that there might indeed be some
fundamental lattice at a scale below current observations.  My
qualms are that this opens up a vast number of
variations.  In the past the criterion of renormalizability has proven
quite useful in limiting the theories used in particle physics.
We have no strong evidence, apart from possibly gravity, that nature uses
any other fundamental interactions.  With a fundamental lattice this
need for renormalizability becomes obscured.

Recently there has been considerable interest in fermionic theories based
on fundamental four fermion couplings.  See for example
Ref.~[\refname\kahanaref].  While non-renormalizable, these
can give rise to interesting field theoretic phenomena through dynamical
symmetry breaking.  It is conceivable that lattice methods
may be of use here.  In such theories, a natural cutoff scale would
be the Planck length.  Ref.~[\refname\xueref] has suggested that
such a theory might circumvent the fermion doubling problems of
conventional approach.

Finally, let me note that after twenty years, lattice
gauge theory remains a thriving industry.  The method is still the
most viable approach to study non-perturbative
phenomena in quantum field theory.
Despite these successes, the fundamental problems
with lattice fermions show that we still have an acute need for new ideas.

\bigskip
\noindent References
\medskip
\item{\wilsonref.} K. Wilson, Phys. Rev. D10, 2445 (1974).
\item{\latticeref.} Nucl. Phys. B Proceedings Supplements, vols
4, 9, 17, 20, 26, and 30.
\item{\fnalref.} A. El-Khadra, G. Hockney, A. Kronfeld, and P. Mackenzie,
Phys. Rev. Lett. 69, 729 (1992).
\item{\weingartenref.} F. Butler, H. Chen, J. Sexton, A. Vaccarino, and
D. Weingarten, Phys. Rev. Lett. 70, 2849 (1993).
\item{\gavairef.} For a review see R. Gavai, in {\sl Quantum Fields on
the Computer}, M. Creutz, ed., (World Scientific, 1992) pp. 51-124.
\item{\bernardsoniref.} C. Bernard and A. Soni, in {\sl Quantum Fields on
the Computer}, M. Creutz, ed., (World Scientific, 1992) pp. 362-427.
\item {\blsref.} C. Bernard, J. Labrenz, and A. Soni,
preprint UW-PT-93-06 (1993).
\item {\kaplanref.} D. Kaplan, Phys. Lett. B288,
342 (1992); M. Golterman, K. Jansen, D. Kaplan,
Phys. Lett. B301, 219 (1993).
\item {\slavnovref.} S. Frolov and A. Slavnov, Max-Planck Inst.
preprint MPI-Ph 93-12 (1993).
\item {\neubergerref.} R. Narayanan and H. Neuberger,
Phys. Lett. B302, 62 (1993); preprint RU-93-34.
\item {\aokigockschref.} S. Aoki and A. Gocksch, Phys. Rev. D45, 3845 (1992)
and references therein.
\item {\jansenschmaltzref.} Karl Jansen and Martin Schmaltz, Phys. Lett.
B296, 374 (1992).
\item {\callanharveyref.} C. Callan and J. Harvey, Nucl. Phys. B250, 427
(1985).
\item {\thooftref.} G. t'Hooft, Phys. Rev. Lett. 37, 8 (1976); Phys. Rev. D14,
3432 (1976).
\item {\gravityref.} J. Jurkiewicz, Nucl. Phys. B (Proc. Suppl.) 30, 108
(1993).
\item {\kahanaref.} D. Kahana and S. Kahana, Phys. Rev. D43, 2361 (1991).
\item {\xueref.} G. Preparata and S.-S. Xue, Univ. of Milan preprint MITH 93/5
(1993).

\vfill\eject
\bye